\documentclass[aps,prd,nofootinbib,showpacs,twocolumn,floatfix]{revtex4}
\usepackage{amssymb,amsfonts}
\usepackage{bm} % bold math
\usepackage{graphicx}
\newcommand{\w}[1]{\bm{#1}}
\newcommand{\be}{\begin{equation}}
\newcommand{\ee}{\end{equation}}
\newcommand{\bea}{\begin{eqnarray}}
\newcommand{\eea}{\end{eqnarray}}
\newcommand{\Hor}{{\mathcal H}}
\newcommand{\M}{{\mathcal M}}
\newcommand{\Sp}{{\mathcal S}}

\newcommand{\el}{\w{\ell}}

\newcommand{\Lie}[1]{\bm{\mathcal L}_{\w{#1}}\,}

\newcommand{\DS}{\w{\mathcal{D}}}

\newcommand{\volS}{\w{\epsilon}_{\scriptscriptstyle\mathcal{S}}}

\begin{document}

\title{Area evolution, bulk viscosity and entropy principles for dynamical horizons}

\author{Eric Gourgoulhon}
\email[]{eric.gourgoulhon@obspm.fr} 
\author{Jos\'e Luis Jaramillo}
\email[]{jose-luis.jaramillo@obspm.fr} 
\affiliation{Laboratoire de
l'Univers et de ses Th\'eories, UMR 8102 du C.N.R.S., Observatoire
de Paris, F-92195 Meudon Cedex, France}

\date{6 September 2006}

\begin{abstract}
We derive from Einstein equation an evolution law for the area of a trapping or dynamical horizon. The solutions to this differential equation show a causal behavior. Moreover, in a viscous fluid analogy, the equation can be interpreted as an energy balance law, yielding to a positive bulk viscosity. These two features contrast with the event horizon case, where the non-causal evolution of the area and the negative bulk viscosity require teleological boundary conditions. This reflects the local character of trapping horizons as opposed to event horizons. Interpreting the area as the entropy, we propose to use an area/entropy evolution principle to select a unique dynamical horizon and time slicing in the Cauchy evolution of an initial marginally trapped surface.
\end{abstract}


\maketitle

%%%%%%%%%%%%%%%%%%%%%%%%%%%%%%%%%%%%%%%%%%%%%%%%%%%%%%%%%%%%%%%%%%%%%%%%%%%%%%%

\noindent\emph{Introduction.}
A new paradigm has recently emerged in the theoretical approaches to black holes,
following the introduction of \emph{future outer trapping horizons (FOTH)} by 
Hayward \cite{Haywa94b,Haywa04c-b} and that of \emph{dynamical horizons (DH)}
by Ashtekar and Krishnan \cite{AshteK03}
(see Refs.~\cite{AshteK04,Booth05} for reviews). This new approach relies
on local characterizations of black holes, via trapped surfaces, 
and contrasts with the traditional treatment which is 
based on the notion of \emph{event horizon (EH)}. The latter is a highly
non-local concept and
requires the knowledge of the whole spacetime to be determined. 
This feature makes the EH a not very
practical representation of black holes for studies 
in numerical relativity and quantum gravity, and 
motivated the new approach. 

In this note, we focus on the evolution of the area of cross sections
of FOTHs and DHs. We obtain an area law of different nature from the ``area increase law'' 
derived by Ashtekar and Krishnan \cite{AshteK03,AshteK04}. 
Inspired by the membrane paradigm developed for
EHs \cite{Damou79-82,ThornPM86-PT86}, we treat the cross sections of FOTHs
and DHs as viscous fluid bubbles and interpret the area law as an internal energy balance equation, completing the Navier-Stokes-like momentum law 
derived previously \cite{Gourg05}. It is then apparent that the bulk viscosity
is positive. Associating area with entropy, we also discuss the use of an 
entropy principle to pick up  a unique dynamical horizon in an evolution. 
Both a maximum entropy production and a time convexity requirement (interpretable in terms
of Clausius-Duhem inequality) are considered.

%%%%%%%%%%%%%%%%%%%%%%%%%%%%%%%%%%%%%%%%%%%%%%%%%%%%%%%%%%%%%%%%%%%%%%%%%%%%%%%

\noindent\emph{Evolution of the horizon area.}
Both the EH and the FOTH/DH approaches to black holes
can be formulated in terms of a hypersurface $\Hor$ embedded in 
a four-dimensional spacetime $(\M,\w{g})$, and foliated by a family of closed (topologically
$\mathbb{S}^2$) spacelike 2-surfaces $(\Sp_t)_{t\in\mathbb{R}}$.
If it represents an EH, $\Hor$ is a null hypersurface and any foliation 
$(\Sp_t)_{t\in\mathbb{R}}$ is admissible.
In the case of a DH, $\Hor$ is a spacelike hypersurface and
the foliation $(\Sp_t)_{t\in\mathbb{R}}$ is unique [$t$ defined
up to a relabeling
$t\mapsto t'=f(t)$], each $\Sp_t$ being a marginally trapped 
surface \cite{AshteG05}. 

Given the foliation $(\Sp_t)_{t\in\mathbb{R}}$ of $\Hor$, there is a unique
``time evolution'' vector $\w{h}$ that is tangent to 
$\Hor$, orthogonal to $\Sp_t$ and obeys $\Lie{h}t=1$, where $\Lie{h}$ denotes
the Lie derivative along $\w{h}$: $\Lie{h}t=h^\mu\partial_\mu t$.
The latter property implies that the 2-surfaces $\Sp_t$ are Lie dragged by $\w{h}$.
Let $C$ be half the scalar square of $\w{h}$ with respect to the metric $\w{g}$:
$\w{h}\cdot\w{h}=2C$. It is easy to see that the sign of $C$ gives the signature
of the hypersurface $\Hor$: $C$ is positive, zero and negative for respectively 
spacelike, null and timelike hypersurfaces. 
There exists a unique pair $(\el,\w{k})$ of null vectors normal to $\Sp_t$ and
a unique vector $\w{m}$ normal to $\Hor$ such that
\be \label{e:def_l_k_m}
    \w{h} = \el - C \w{k},\qquad  
    \w{m} = \el + C \w{k}\quad\mbox{and}\quad
    \el\cdot\w{k} = - 1 .
\ee
For any vector field $\w{v}$ normal to $\Sp_t$, such as $\w{h}$, $\w{m}$, $\el$ or $\w{k}$, we define the \emph{expansion} $\theta^{(\w{v})}$ and the 
\emph{shear tensor} 
$\w{\sigma}^{(\w{v})}$ of the surface $\Sp_t$ when Lie-dragged along $\w{v}$
by 
\be \label{e:def_exp_shear}
   \Lie{v} \w{q} = \theta^{(\w{v})} \w{q} + 2 \w{\sigma}^{(\w{v})} 
    \quad\mbox{and}\quad
     \mathrm{tr}\, \w{\sigma}^{(\w{v})} = 0 ,
\ee
where $\w{q}$ is the induced metric on $\Sp_t$ ($\w{q}$ is positive definite
since $\Sp_t$ is assumed to be spacelike), $\Lie{v} \w{q}$ is its Lie derivative
resulting from the dragging of the surface $\Sp_t$ along the normal vector $\w{v}$,
and $\mathrm{tr}\, \w{\sigma}^{(\w{v})}$ is the trace of $\w{\sigma}^{(\w{v})}$
with respect to the metric $\w{q}$. 
A consequence of (\ref{e:def_exp_shear}) is that $\theta^{(\w{v})}$ governs
the variation of the surface element 2-form $\volS$ of $\Sp_t$
according to $\Lie{v} \volS = \theta^{(\w{v})} \volS$.

Let us recall that $\Sp_t$ is called a \emph{trapped surface}  if $\theta^{(\w{k})}<0$ and 
$\theta^{(\el)}<0$, and 
\emph{marginally trapped surface (MTS)} if $\theta^{(\w{k})}<0$ and 
$\theta^{(\el)}=0$. The hypersurface 
$\Hor=\bigcup_{t\in\mathbb{R}} \Sp_t$ is said to be
a \emph{future outer trapping horizon (FOTH)} if (i) each $\Sp_t$ is a 
MTS and (ii) the (local) outermost condition $\Lie{k}\theta^{(\el)}<0$ is satisfied \cite{Haywa94b}. $\Hor$ is said to be a \emph{dynamical horizon (DH)} if (i) each $\Sp_t$ is a MTS and (ii) $\Hor$ is a spacelike hypersurface
\cite{AshteK03}. Note that in generic dynamical situations, the notions
of FOTH and DH are equivalent \cite{Booth05}. In stationary
situations, a FOTH becomes a null hypersurface, whereas a DH (which by definition is spacelike) cannot exist; it should be replaced by the notion of 
\emph{isolated horizon (IH)} \cite{AshteBF99,AshteK04,Booth05,GourgJ06a}.

If $\Hor$ is an EH, the 2-surfaces $\Sp_t$ are not MTS, except in stationary configurations (Kerr black hole). On the contrary
they are expanding, by the famous Hawking area increase law \cite{Hawki72}:
$\theta^{(\el)}>0$. 

Let us denote by $\kappa$ the component along $\el$ of the ``acceleration'' of $\w{h}$ in the decomposition \cite{GourgJ06c}
\be \label{e:def_kappa}
	\w{\nabla}_{\w{h}} \w{h} = \kappa \, \el + (C\kappa - \Lie{h} C) \w{k}
	- \DS C , 
\ee
where $\w{\nabla}$ is the spacetime connection and
$\DS$ the connection associated with the metric $\w{q}$ in $\Sp_t$. 
Besides let $\w{\Omega}^{(\el)}$ be the \emph{normal fundamental form} of the surface $\Sp_t$ (also called \emph{connection on the normal bundle}) defined by
$\w{\Omega}^{(\el)} := - \w{k}\cdot \w{\nabla}_{\vec{\w{q}}} \, \el$, where
$\vec{\w{q}}$ denotes the orthogonal projector on the surface $\Sp_t$
(see e.g. Sec.~III.D of Ref.~\cite{Gourg05}). 

From the Einstein equation, one can derive the following evolution law
for any foliated hypersurface $\Hor$ (details are provided in 
Ref.~\cite{GourgJ06c}):
\begin{widetext}
\be \label{e:Lh_theta_m}
\Lie{\w{h}} \theta^{(\w{m})} =  \kappa \, \theta^{(\w{h})}
	- \frac{1}{2} \theta^{(\w{h})} \theta^{(\w{m})}
	- \w{\sigma}^{(\w{h})} : \w{\sigma}^{(\w{m})} 
	- 8\pi \w{T}(\w{m},\w{h}) 
	+ \theta^{(\w{k})} \Lie{\w{h}} C
	+ \DS\cdot\left( 2 C \vec{\w{\Omega}}^{(\el)} - \vec{\DS} C \right) ,
\ee
where $\w{T}$ is the energy-momentum tensor of matter (if any), 
an upper arrow indicates index raising with the metric $\w{q}$
and the notation ``:'' stands for the double contraction, i.e. 
$\w{\sigma}^{(\w{h})} : \w{\sigma}^{(\w{m})} := \sigma^{(\w{h})}_{ab}
	\sigma^{(\w{m})ab}$.
If we specialize Eq.~(\ref{e:Lh_theta_m}) to the cases of (i) an EH 
and (ii) a FOTH or a DH, we obtain respectively
\begin{eqnarray} 
\Lie{\el} \theta^{(\el)} + (\theta^{(\el)})^2 - \kappa \, \theta^{(\el)} & = & 
	\frac{1}{2} (\theta^{(\el)})^2 
	- \w{\sigma}^{(\el)} : \w{\sigma}^{(\el)} 
	- 8\pi \w{T}(\el,\el) \label{e:evol_th_EH} \\
\Lie{\w{h}} \theta^{(\w{h})} 
	+ (\theta^{(\w{h})})^2 + \kappa \, \theta^{(\w{h})}   & = &  
	\frac{1}{2} (\theta^{(\w{h})})^2 
	+ \w{\sigma}^{(\w{h})} : \w{\sigma}^{(\w{m})} 
	+ 8\pi \w{T}(\w{m},\w{h}) 
	- \theta^{(\w{k})} \Lie{h} C 
	+ \DS\cdot\left( \vec{\DS} C  - 2 C \vec{\w{\Omega}}^{(\el)} \right) .
	 \label{e:evol_th_FOTH}
\end{eqnarray}
\end{widetext}
For the EH, we have used the null character of $\Hor$, which implies $C=0$ and $\w{h}=\w{m}=\el$, yielding Eq.~(\ref{e:evol_th_EH}). This is nothing but
the null Raychaudhuri equation for a surface-orthogonal congruence
\cite{HawkiH72}. In this case
Eq.~(\ref{e:def_kappa}) reduces to $\w{\nabla}_{\el}\,  \el = \kappa \el$, 
i.e. $\kappa$ is the non-affinity coefficient of $\el$ and coincides with the so-called \emph{surface gravity} for a stationary horizon, provided that $\el$
is normalized in terms of the stationarity Killing vector.  
For the FOTH/DH case [Eq.~(\ref{e:evol_th_FOTH})], we have used the property $\theta^{(\el)}=0$ to write $\theta^{(\w{m})}=-\theta^{(\w{h})}$, since 
Eq.~(\ref{e:def_l_k_m}) gives
$\theta^{(\w{m})}=-\theta^{(\w{h})}+2\theta^{(\el)}$. 

The area of the 2-surface $\Sp_t$ is $A(t)=\int_{\Sp_t} \volS$.
Since the surfaces $\Sp_t$ are Lie dragged by the vector $\w{h}$
associated with $t$, the first and second derivative of the area 
with respect to $t$ are
$dA/dt = \int_{\Sp_t} \theta^{(\w{h})}\volS$ and 
$d^2A/dt^2 =  \int_{\Sp_t} \left[ \Lie{h} \theta^{(\w{h})} 
+ (\theta^{(\w{h})})^2 \right] \volS$.
Then, assuming $C>0$ in Eq.~(\ref{e:evol_th_FOTH}) (i.e. considering a DH only), 
we may introduce
$\kappa' := -C^{-1} \el\cdot\w{\nabla}_{\w{h}} \w{h} = \kappa - \Lie{h}\ln C$
[cf. Eq.~(\ref{e:def_kappa})],
and integrate Eqs.~(\ref{e:evol_th_EH}) and (\ref{e:evol_th_FOTH})
over $\Sp_t$, noticing that the integral of the divergence term in 
Eq.~(\ref{e:evol_th_FOTH}) vanishes, to get respectively
\bea
   \frac{d^2 A}{dt^2} - \bar\kappa \frac{dA}{dt} &= & - \int_{\Sp_t} \Big[   
	8\pi \w{T}(\el,\el)  + \w{\sigma}^{(\el)} \!:\!\w{\sigma}^{(\el)} 
	  \nonumber \\
	& & 
 \qquad  - \frac{(\theta^{(\el)})^2}{2} 
  + (\bar\kappa-\kappa) \theta^{(\el)} \; \Big] \, \volS  , \label{e:evol_A_EH}
\eea
\bea
	\frac{d^2 A}{dt^2} + \bar\kappa' \frac{dA}{dt} & = &
	\int_{\Sp_t}  \Big[  
	8\pi \w{T}(\w{m},\w{h})  
	+ \w{\sigma}^{(\w{h})} \!:\!\w{\sigma}^{(\w{m})} 
	 \nonumber \\
	& & 
\quad + \frac{(\theta^{(\w{h})})^2}{2}
+ (\bar\kappa'-\kappa') \theta^{(\w{h})}  \; \Big] \, \volS , \label{e:evol_A_DH}
\eea
where $\bar\kappa$ and $\bar\kappa'$ denote the mean value over $\Sp_t$
of $\kappa$ and $\kappa'$: $\bar\kappa = \bar\kappa(t) := A^{-1} \int_{\Sp_t} \kappa\volS$ (idem for $\kappa'$).
Assume for a moment that $\bar\kappa$ and $\bar\kappa'$ are constant and positive, as for the EH of a Kerr black hole with foliations compatible
with the stationarity Killing vector. 
Let us consider first the evolution of an EH, i.e. the differential
equation (\ref{e:evol_A_EH}).  The general solution of the 
corresponding homogeneous equation is $A(t)=\mathrm{const}_1\cdot\exp(\bar\kappa t)+\mathrm{const}_2$. 
Thus, if one were solving Eq.~(\ref{e:evol_A_EH}) as a Cauchy problem, one would obtain exponentially diverging solutions (since $\bar\kappa>0$). 
It is well known that the correct treatment must be \emph{teleological} 
(\cite{HawkiH72,Damou79-82} or Sec. VI.C.6 of Ref.~\cite{ThornPM86-PT86}), i.e. one imposes
the boundary condition $dA/dt=0$ at $t=+\infty$, to get 
\be \label{e:dAdt_EH}
	\frac{dA}{dt} = \int_t^{+\infty} D(u) e^{\bar\kappa(t-u)} \, du ,
\ee
where $D=D(t)$ stands for the integral in the right-hand side of
Eq.~(\ref{e:evol_A_EH}).
On the contrary, the differential equation (\ref{e:evol_A_DH})
for the DH area is such that the solutions of the homogeneous
equation are decaying exponentially. Accordingly, the treatment as a standard Cauchy
problem from the initial condition $dA/dt(0)={\dot A}_0$ leads to the
non-diverging solution
\be \label{e:dAdt_DH}
	\frac{dA}{dt} = {\dot A}_0 + \int_0^t D'(u) e^{\bar\kappa'(u-t)} \, du ,
\ee
where $D'=D'(t)$ stands for the integral in the right-hand side of
Eq.~(\ref{e:evol_A_DH}).
The striking difference between Eqs.~(\ref{e:dAdt_EH}) and (\ref{e:dAdt_DH})
is that Eq.~(\ref{e:dAdt_DH}) is causal (the solution at a given instant $t$
depends only on the behavior of the source $D'$ at instants $u\leq t$),
whereas Eq.~(\ref{e:dAdt_EH}) is not. This reflects the non-local 
character of EHs 
mentioned in the Introduction. Of course, in general $\bar\kappa$ and $\bar\kappa'$
are not constant 
(except for small perturbations of a stationary black hole \cite{BoothF04}),
but the behavior described above should remain the same. 

Comparing with previous works, one can show that at the limit of small
departure from an IH, the area law (\ref{e:evol_A_DH}) reduces to that
established for slowly evolving horizons by Booth \& Fairhurst \cite{BoothF04}. 
In the full dynamical regime, the evolution law (\ref{e:evol_A_DH}) is
different from the area law obtained by Ashtekar \& Krishnan \cite{AshteK03,AshteK04}
(see also \cite{Haywa06}).
Indeed the latter is derived from a different component of Einstein 
equation: the $\w{T}(\w{m},\el)$ one instead of $\w{T}(\w{m},\w{h})$
for Eq.~(\ref{e:evol_A_DH}).
More precisely, the Ashtekar-Krishnan law is written
in terms of the areal radius $R:=\sqrt{A/4\pi}$  
and expresses the variation of $R$ between two
surfaces $\Sp_{t_1}$ and $\Sp_{t_2}$ (cf. Eq.~(3.25) of Ref.~\cite{AshteK03}). Using our notations (cf. Table~II of Ref.~\cite{Gourg05} for the correspondence) its differential version (i.e. writing $t_2=t_1+dt$) is
\bea
\frac{1}{2} \frac{dR}{dt}  &=& \frac{1}{8\pi} \int_{\Sp_t}
	\left[ \w{\Omega}^{(\w{\tilde\ell})} \cdot
	\vec{\w{\Omega}}^{(\w{\tilde\ell})}
	+ \frac{1}{C} \w{\sigma}^{(\el)}\!:\! \w{\sigma}^{(\el)} 
	\right] \frac{dR}{dt} \, \volS \nonumber \\
	& & + \int_{\Sp_t} \frac{1}{C} \w{T}(\w{m},\el) \,  
	\frac{dR}{dt} \, \volS ,
		\label{e:AK_law}	
\eea
where $\w{\Omega}^{(\w{\tilde\ell})}  := \w{\Omega}^{(\el)} - \DS\ln C$.
Notice that this is a first order equation in $A(t)=4\pi R(t)^2$, whereas
Eq.~(\ref{e:evol_A_DH}) is of second order. Moreover,  
it contains $dR/dt$ on both sides, so that one can divide 
by $dR/dt$ to get an equation which does not contain
$R(t)$. In this respect, Eq.~(\ref{e:AK_law}) does not appear as 
an evolution equation for $A(t)$. Actually, as shown by Hayward \cite{Haywa04c-b}, 
Eq.~(\ref{e:AK_law}) can be obtained by integrating over $\Sp_t$ the relation expressing
that $\Lie{h}\theta^{(\el)}=0$ on $\Hor$. This relation
involves $\w{T}(\w{m},\el)$, in the same manner as Eq.~(\ref{e:Lh_theta_m}) relates
$\Lie{\w{h}} \theta^{(\w{m})}$ and $\w{T}(\w{m},\w{h})$. Explicitly we have \cite{Haywa04c-b,GourgJ06c} 
\bea
  & & C \left[ \DS\cdot \vec{\w{\Omega}}^{(\w{\tilde\ell})}
	+ \w{\Omega}^{(\w{\tilde\ell})} \cdot \vec{\w{\Omega}}^{(\w{\tilde\ell})}
	- \frac{1}{2} \mathcal{R} \right] 
	+ \w{\sigma}^{(\el)}\!:\!\w{\sigma}^{(\el)} \nonumber \\
	& & \ \qquad \qquad  + 8\pi \w{T}(\w{m},\el) = 0 , \label{e:Tml}
\eea
where $\mathcal{R}$ denotes the Ricci scalar associated with the metric $\w{q}$
in $\Sp_t$. Dividing Eq.~(\ref{e:Tml}) by $8\pi C$, integrating over $\Sp_t$, invoking the Gauss-Bonnet theorem to set the integral of $\mathcal{R}$
to $8\pi$ and multiplying by $dR/dt$ leads to the differential form
(\ref{e:AK_law}) of Ashtekar-Krishnan law. 

%%%%%%%%%%%%%%%%%%%%%%%%%%%%%%%%%%%%%%%%%%%%%%%%%%%%%%%%%%%%%%%%%%%%%%%%%%%%%%%

\noindent\emph{Energy dissipation and bulk viscosity.}
In the membrane paradigm approach to black holes,
Price and Thorne \cite{ThornPM86-PT86} defined the \emph{surface energy
density} of an EH as $\varepsilon := -\theta^{(\el)}/8\pi$ and
interpreted Eq.~(\ref{e:evol_th_EH}) as an energy balance law, 
with heat production resulting from viscous stresses. 
By analogy, let us define the \emph{surface energy density} of a FOTH/DH
as $\varepsilon := -\theta^{(\w{m})}/8\pi$, where the role of the normal to $\Hor$ is now taken by $\w{m}$ instead of $\el$. Since 
$\theta^{(\w{m})} = -\theta^{(\w{h})}$ for a FOTH/DH, we have
$\varepsilon = \theta^{(\w{h})}/8\pi$ and we may rewrite 
Eq.~(\ref{e:evol_th_FOTH}) as
\bea
	\Lie{h}\varepsilon + \theta^{(\w{h})} \varepsilon & = & 
	- \frac{\kappa}{8\pi} \theta^{(\w{h})} 
	+  \frac{1}{8\pi} \w{\sigma}^{(\w{h})} : \w{\sigma}^{(\w{m})}
	+ \frac{(\theta^{(\w{h})})^2}{16\pi}   \nonumber \\
	& & 
	- \DS\cdot\w{Q} 
	+ \w{T}(\w{m},\w{h})  
	- \frac{\theta^{(\w{k})}}{8\pi} \Lie{h} C ,  \label{e:cons_ener}
\eea
with $\w{Q} := \frac{1}{4\pi} \left[ C \vec{\w{\Omega}}^{(\el)} - 1/2\, \vec{\DS} C  \right] = - \frac{C}{4\pi} \vec{\w{\varpi}}$, where $\w{\varpi}$ is the
\emph{anholonomicity 1-form} (or \emph{twist 1-form}) of the 2-surface
$\Sp_t$ \cite{Haywa94b} (see also Sec.~IV.A of Ref.~\cite{Gourg05}) and
$\vec{\w{\varpi}}$ denotes its vector dual.
It is worth to write aside Eq.~(\ref{e:cons_ener}) the generalized 
Damour-Navier-Stokes equation \cite{Gourg05}
\bea
	\Lie{h} \w{\pi} 
	+ \theta^{(\w{h})} \w{\pi} & = & - \DS  \left(\frac{\kappa}{8\pi} \right)
	+ \DS\cdot \left( 
	\frac{\vec{\w{\sigma}}^{(\w{m})}}{8\pi} \right)
	+ \frac{\DS \theta^{(\w{h})}}{16\pi}  \nonumber \\
	& &  - \w{T}(\w{m},\vec{\w{q}}) 
	+ \frac{\theta^{(\w{k})}}{8\pi} \DS C ,  \label{e:DNS}
\eea
where $\w{\pi} := - \w{\Omega}^{\el} / 8\pi$.
It is striking that Eqs.~(\ref{e:cons_ener})-(\ref{e:DNS}) are fully analogous to
the equations that govern a two-dimensional 
non-relativistic fluid of internal energy density $\varepsilon$, momentum density 
$\w{\pi}$, pressure $\kappa/8\pi$, 
shear stress tensor $\w{\sigma}^{(\w{m})}/8\pi$,
bulk viscosity $\zeta = 1/16\pi$, shear strain tensor $\w{\sigma}^{(\w{h})}$, expansion $\theta^{(\w{h})}$, subject to the external force density
$- \w{T}(\w{m},\vec{\w{q}}) + \theta^{(\w{k})}/8\pi\,  \DS C$, external energy production rate
$\w{T}(\w{m},\w{h})- \theta^{(\w{k})}/8\pi\, \Lie{h} C $ and  heat flux $\w{Q}$
(see e.g. Ref.~\cite{Rieut97}). 
Let us notice that the shear viscosity $\mu$ does not appear in 
Eqs.~(\ref{e:cons_ener})-(\ref{e:DNS}), because the standard 
Newtonian-fluid relation between the shear stress tensor $\w{\sigma}^{(\w{m})}/8\pi$ and the shear strain tensor $\w{\sigma}^{(\w{h})}$, namely $\w{\sigma}^{(\w{m})}/8\pi = 2 \mu \w{\sigma}^{(\w{h})}$, does not hold.
Here we have 
$\w{\sigma}^{(\w{m})}/8\pi = 
[ \w{\sigma}^{(\w{h})} + 2 C \w{\sigma}^{(\w{k})}]/8\pi$, so that the 
Newtonian-fluid assumption is fulfilled only if $C=0$ (IH limit). 
On the contrary, it appears from Eqs.~(\ref{e:cons_ener})-(\ref{e:DNS}) that 
the trace part of the viscous stress tensor $\w{S}_{\rm visc}$ obeys the 
Newtonian-fluid law, being proportional to the 
trace part of the strain tensor (i.e. the expansion $\theta^{(\w{h})}$): 
${\rm tr} \, \w{S}_{\rm visc} = 3 \zeta \theta^{(\w{h})}$.  

Let us point out two differences with the EH case 
\cite{Damou79-82,ThornPM86-PT86}. First the heat flux $\w{Q}$ is not 
vanishing for a FOTH/DH, whereas it was zero for an EH.
Notice that $\w{Q}$ is a vector tangent to $\Sp_t$ so that the integration 
of Eq.~(\ref{e:cons_ener}) over the closed surface $\Sp_t$ to get a global
internal energy balance law would not contain any net heat flux. 
The second major difference is that the bulk viscosity
$\zeta$ is positive, being equal to $1/16\pi$, 
whereas it was found to be negative, being equal to $-1/16\pi$, for 
EHs \cite{Damou79-82,ThornPM86-PT86}. 
This negative value, which would yield to a dilation or contraction instability
in an ordinary fluid, is in agreement with the tendency of a null hypersurface
to continually contract or expand, the EH 
being stabilized by the teleological condition imposing its expansion to vanish
in the far future. 
The positive value of the bulk viscosity found here shows that 
FOTHs and DHs behave as ``ordinary''
physical objects and is in perfect agreement with their local nature. 

%%%%%%%%%%%%%%%%%%%%%%%%%%%%%%%%%%%%%%%%%%%%%%%%%%%%%%%%%%%%%%%%%%%%%%%%%%%%%%%

\noindent\emph{Entropy evolution and choice of $\Hor$.} 
Hitherto we have discussed the horizon $\Hor$ as a given hypersurface in $\M$.
We adopt now a 3+1 approach and consider, in the line of Ref.~\cite{AnderMS05},
the world-tube evolution of a MTS $\Sp_0$ contained in an initial
Cauchy slice $\Sigma_0$ of a  3+1 foliation 
$(\Sigma_t)_{t\in\mathbb{R}}$. Denoting by $\w{s}$ the unit-normal vector to $\Sp_0$ contained 
in $\Sigma_0$ and by $\w{n}$ the unit-normal vector to  $\Sigma_0$, 
we decompose $\Hor$'s ``time evolution'' vector as $\w{h} = N \w{n} + b \w{s}$, where $N$ is the
lapse function associated with the time parameter $t$ of the 3+1 foliation. 
Introducing the null vectors $\hat{\el}:= \w{n}+\w{s}$ and $\hat{\w{k}}:= (\w{n}-\w{s})/2$,
we get  $\w{h} = (b+N)/2\; \hat{\el} + (b-N)\hat{\w{k}}$ and
$C=(b^2-N^2)/2$.
Combined uniqueness results from Refs.~\cite{AshteG05,AnderMS05} show that:
(i) to each 3+1 foliation $(\Sigma_t)_{t\in\mathbb{R}}$ there corresponds a unique DH containing  $\Sp_0$ and sliced by MTSs
$\Sp_t\subset\Sigma_t$; 
and (ii) different
3+1 slicings lead generically to different DH. 
In other words, the evolution of  $\Sp_0$ into MTSs 
is an ill-defined concept, since different DHs pass
through $\Sp_0$. A natural question consists in introducing a notion of preferred 
DH or, equivalently, of a preferred 3+1 slicing.

With this ultimate aim, let us push forward the viscous fluid interpretation of 
FOTH/DHs. As an ever growing quantity on $\Hor$, it is natural to interpret the
area $A$ as proportional to the entropy of the horizon, the equilibrium case providing
the relation $S= \frac{k_B}{4\hbar} A$ \cite{Haywa94b,AshteK03,BoothF04}.
In this sense, Eq.~(\ref{e:evol_th_EH}) for an EH has been interpreted in 
the membrane paradigm \cite{Damou79-82,ThornPM86-PT86} as a viscous fluid 
entropy dissipation equation for the differential element of
entropy, with the specificity of involving a
second time derivative term in the entropy production. 
Instead of the analogous discussion for FOTH/DHs 
based on Eq.~(\ref{e:evol_th_FOTH}), 
we take the condition $\Lie{h}\theta^{(\el)}=0$, i.e. Eq.~(\ref{e:Tml}), as
a starting point. When expressed in terms of 3+1 quantities,
it becomes an elliptic equation on $(b-N)$ with a source
proportional to $(b+N)$ \cite{Eardl98,GourgJ06c}.  In this 3+1 approach, a second
relation between $(b-N)$ and $(b+N)$ is necessary in order to determine $\Hor$. 

In a first attempt to get this second relation, 
we consider a {\it maximum entropy production} criterion, which can be motivated in the context of non-equilibrium thermodynamics \cite{Greve03}.
Maximizing $dS/dt$ leads to \cite{GourgJ06c}
\be
b-N = -\mathrm{const}\cdot\theta^{(\hat{\w{k}})}  . \label{e:max_dSdt}
\ee
In loose terms, and independently of thermodynamic considerations, 
Eq.~(\ref{e:max_dSdt}) singles out the DH that approaches ``the fastest'' to the EH.
However it leads to a function $A(t)$ which is only
${\cal C}^0$ in the matching with an initial IH. 

Alternatively, we can control the response of $\Hor$ to the
arrival of energy/matter, by prescribing the convexity of the entropy in time. We propose a phenomenological choice for the second derivative of the area element 
$\ddot{a}:=\Lie{h} \theta^{(\w{h})} + (\theta^{(\w{h})})^2$:
\be
\ddot{a} = F[\mathrm{sources}] + \alpha \frac{\theta^{(\w{h})}}{C} \Lie{h}C 
+ \beta  \frac{\theta^{(\w{h})}}{C} \Lie{h}\Lie{h}C , \label{e:convex_a} 
\ee
such that $F[\mathrm{sources}]>0$ whenever matter or gravitational radiation
crosses the horizon [e.g. $\w{T}(\el,\el)\neq 0$ or
$\w{\sigma}^{(\el)}\neq 0$], and $F[\mathrm{sources}]\leq0$ otherwise. 
Inserting Eq.~(\ref{e:convex_a}) into Eq.~(\ref{e:evol_th_FOTH}) leads
to a second-order evolution
equation for $C=(b^2-N^2)/2$, whose resolution provides the additional
relation between $(b-N)$ and $(b+N)$.
Initial condition $C=0$ guarantees the  ${\cal C}^1$
matching with an IH, whereas the choice of initial $\Lie{h}C$ amounts
to the choice of DH (consistency requires an appropriate choice of
parameters $\alpha$, $\beta$). Alternatively, the equation for $C$
can be seen as a (now first-order) balance equation for the element of
entropy.
Interpreting it as a non-equilibrium thermodynamics Clausius-Duhem-like inequality \cite{Greve03}
(by enforcing the positivity of the entropy production),
provides a guideline for fixing the phenomenological
term $F[\mathrm{sources}]$.
\noindent Other phenomenological options to Eq. (\ref{e:convex_a}) can
be proposed, and we aim at exploring them numerically \cite{GourgJ06c}. 
A tempting possibility is to base
the (unavoidable) choice of DH, upon an entropy principle derived solely from the 
structure of the hyperbolic system defined by (part of ) the Einstein
equations on $\Hor$.

\noindent{\em Acknowledgements}. We thank J. Metzger and L. Andersson for the discussion on the maximization
on the area change.  JLJ acknowledges the support of the Marie
Curie Intra-European contract MEIF-CT-2003-500885 
within the 6th European Community Framework Programme.

\end{document}